\documentclass[runningheads]{svmult}

\usepackage{makeidx}   
\usepackage{graphicx}  
\usepackage{subeqnar}  
\usepackage{multicol}  
\usepackage{physprbb}  
\makeindex             

\begin{document}
\title*{Torus models for obscuration in type 2 AGN}
\toctitle{Torus models for obscuration in type 2 AGN}
%
%
\titlerunning{Torus models for type 2 AGN}
%
\author{Thomas Beckert\inst{1}
\and Wolfgang J. Duschl\inst{2}
\and Bernd Vollmer\inst{3}
}
\authorrunning{T. Beckert et al.}
%
%
\institute{Max-Planck-Institut f{\"u}r Radioastronomie, 
  Auf dem H{\"u}gel 69, 53121 Bonn, Germany 
  \and Institut f{\"u}r Theoretische Astrophysik 
  der Universit{\"a}t Heidelberg, Tiergartenstra{\ss}e 15, 
  69121 Heidelberg, Germany
  \and CDS, Observatoire astronomique de Strasbourg, UMR 7550, 
  11 rue de l'universit{\'e}, 67000 Strasbourg, France
}

\maketitle              

\begin{abstract}   
  We discuss a clumpy model of obscuring dusty tori 
  around AGN. Cloud-cloud collisions lead to an effective
  viscosity and a geometrically thick accretion disk, 
  which has the required properties of a torus. 
  Accretion in the combined gravitational potential of 
  central black hole and stellar cluster generates free energy, 
  which is dissipated in collisions, and maintains the
  thickness of the torus.   A quantitative
  treatment for the torus in the prototypical Seyfert~2 
  nucleus of NGC~1068 together with a radiative transfer 
  calculation for NIR re-emission from the torus is presented. 
\end{abstract}

%

\section{Introduction: Dusty tori in the unified model of AGN}

Starting with the interpretation of spectropolarimetry of NGC~1068 by  
Antonucci \& Miller \cite{Miller83,Anton85}, 
the difference
between type~1 and type~2 AGN is attributed to
aspect-angle-dependent obscuration.  
The central continuum source is thought to be surrounded by a 
dusty molecular gas distribution,
which is flattened by rotation into a torus or thick disk.
The simplest 
unification scheme assumes that all Seyfert~2 nuclei harbor 
a Seyfert~1 core, so that the ratio of type 1s to 2s measures 
the thickness of the torus.
Present estimates range form 1:4 \cite{Maiolino95}
to 1:1 \cite{Lacy04} when selected in the mid--infrared. 
The torus should therefore have a thickness $H/R \sim 1$.

For more than a decade, clumpy models for the circumnuclear
torus have been discussed \cite{Krolik88}. It was
recognized early on that dust will only survive in distinct clouds,
which are embedded in the inter-cloud medium of the 
geometrically thick torus. This idea was
nonetheless not included in the following radiative transfer
calculations \cite{Pier92,Pier93,Granato94,Efstathiou95} 
of dust emission from these tori.
Only recently, Nenkova et al. \cite{Nenkova02} developed a scheme which
uses the clumpiness of the torus for an approximative and statistical
calculation of the thermal IR-emission from clouds, which are individually
optically thick $\tau_V > 40$. 
This approach resolved some of the problems with earlier
radiative transfer calculations. 
The toroidal distribution of atomic and molecular material 
is also inferred from observed H{\sc i} absorption and H$_2$ emission, 
for example in the nucleus of NGC~4151 \cite{Mundell03}.

In two papers \cite{Vollmer04,Beckert04} we 
studied the consequences of the proposed clumpiness. 
It turns out that a scenario with distinct, quasi-stable clouds, 
which experience frequent cloud-cloud collisions, 
is unavoidable for geometrically thick torus. In this paper we emphasis the 
importance of cloud collisions and present the results of radiative transfer
calculations for the torus in NGC~1068.  

\section{Dusty clouds in the torus}
Cold, molecular and dusty clouds are the basic constituents
of an accretion scenario of the torus. In spite of the complexity 
of their internal physics,
we regard them as spherical clouds, which are sufficiently
described by their radius $r_{\rm Cl}$ and internal sound speed $c_s$.
The clouds must be self-gravitating to prevent dissolving in the 
inter cloud medium,
which implies that the clouds free-fall time and the sound crossing time
are approximately equal. This determines the volume filling 
factor
\begin{equation}
  \phi_V = \frac{32 G \rho_0 r_{\rm Cl}^2}{3 \pi c_s^2}
\end{equation}
of clouds, where $ \rho_0 = \phi_V \rho_{\rm Cl} $ is the 
mean mass density in the torus 
and $\rho_{\rm Cl}$ the density of individual clouds. 
The most important parameter 
of the torus in radiative transfer calculations is the vertical 
optical depth for intercepting a cloud 
\begin{equation}
  \tau = \int {\rm d}z\; \tilde{l}_{\rm coll}^{-1} 
  = \xi \frac{H}{l_{\rm coll}}\;,
\end{equation} 
where $l_{\rm coll}$ is the mean free path in the torus midplane 
($\tilde{l}_{\rm coll}$ is the local value) 
and $\xi$ is the coefficient of the linear relation between $\rho_0$, 
pressure scale height $H$, and surface density $\Sigma = \xi \rho_0 H$, 
which follows from the vertical integration of the density. 
The radius $r_{\rm Cl}$ of clouds can then be expressed as
 a function of $\tau$, $c_s$, and $\Sigma$.

The mean free path is likely to be dominated by the largest clouds present. 
These clouds are more affected by tidal shear than smaller ones, which sets
an upper limit to the possible cloud sizes. We expect that clouds accumulate
at the shear limit when they experience increasing tidal forces 
while being accreted to the center. 
The upper limit is 
\begin{equation}
  r_{\rm Cl} \le \frac{\pi}{\sqrt{8}}\frac{c_s}{\Omega}\;.
\end{equation}  
This is also a lower limit to the surface density 
\begin{equation} \label{eq:Sigma}
  \Sigma \ge \frac{\tau}{\sqrt{8}}\frac{M(R)}{R^2} \frac{c_s}{v_\phi}\,.
\end{equation}
Here $M(R)$ is the total enclosed mass in stars and black hole 
at radius $R$ and $v_\phi$ is the Keplerian circular velocity 
at that position.
Both mass and optical depth will be dominated by the largest clouds 
and the right side of (\ref{eq:Sigma}) will be a good measure of 
the true surface density. This implies a one to one correspondence 
between $\Sigma$ and $\tau$. 
 For the mass of individual clouds
\begin{equation}
  M_{\rm Cl}  \le M(R) \frac{\pi^3}{16 \sqrt{2}}\frac{c_s^3}{v^3_\phi}
\end{equation} 
we find an upper limit, which is much smaller than
the typical mass of molecular clouds in the ISM  of normal galaxies. 
At a radius of $2$\,pc with an enclosed mass of $10^7$\,M$_\odot$ 
and an internal sound speed against gravitational cloud 
collapse of $c_s=2$\,km/s gives a largest possible mass of $35$\,M$_\odot$.
Contrary to the cloud mass we find a lower limit for the 
hydrogen column density through an individual cloud
\begin{equation}
  N_H \ge \frac{M(R)}{\mu m_H} \frac{c_s}{\sqrt{8} R^2 v_\phi}
\end{equation}  
which is about $10^{24}$\,cm$^{-2}$ in the environment close 
to the sublimation radius parameterized above.


For the distribution of clouds in the torus we assume hydrostatic 
equilibrium for the vertical stratification. 
In \cite{Beckert04} we used a modified isothermal 
distribution function of cloud velocities in an external potential, 
which includes a cut-off scale height $x_H$ at which the density 
drops to zero. This leaves room for a wide polar funnel or an
 outflow cavity. We consider only cases, where the cut-off height 
is larger than the pressure scale height $H$. An example for 
the case of NGC~1068 is shown in Fig.~1. The radial structure is 
derived from the stationary accretion scenario described below. 

\section{Cloud collisions and accretion}
The unified model of AGN  requires at least
$\tau \sim 1$ for obscuration of the AGN for line of sights through the torus. 
As $\tau$ is also a dimensionless collision 
frequency $\tau \sim \omega_c/\Omega$, 
this implies that cloud-cloud collisions are frequent in a torus.
For anisotropic velocity dispersions of clouds 
Goldreich \& Tremaine \cite{Goldreich78} derived an effective viscosity 
\begin{equation}\label{eq:visco} 
  \nu = \frac{\tau}{1+\tau^2} \frac{\sigma^2}{\Omega}
\end{equation} 
for angular momentum redistribution\footnote{The enhanced viscosity 
due to collective effects described by Wisdom \& Tremaine \cite{Wisdom88} 
is unimportant for geometrically thick disks or tori with $\Phi_V \ll 1$.}.
The required anisotropy can be determined self-consistently 
for thin accretion disks and we use this limit also for the torus. 

Because in our scenario clouds in a torus are quasi-stable 
and only created or destroyed in cloud collisions, the momentum transfer from 
supernovae and stellar winds is inefficient for maintaining a large 
velocity dispersion of clouds. Accretion in the gravitational potential 
due to cloud collisions generate the free energy to balance 
collisional dissipation.

\begin{figure}
\begin{center}
\includegraphics[width=.8\textwidth]{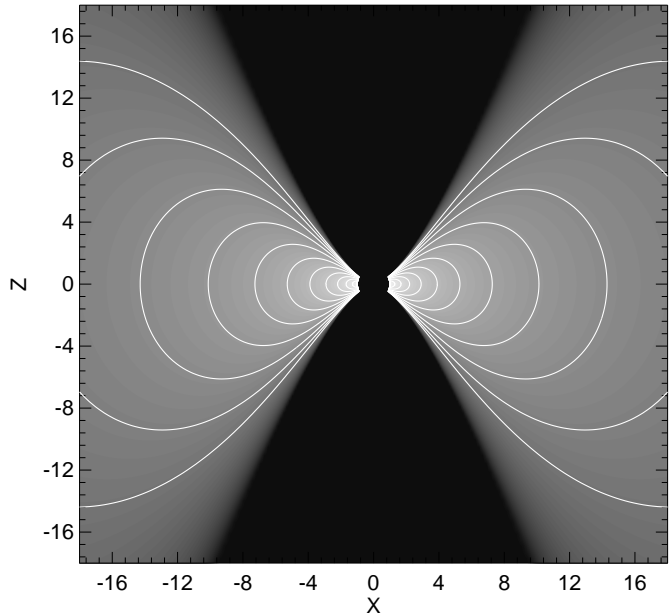}
\end{center}
\caption[]{Meridional cut through the probability density distribution of
    finding a dusty cloud in the torus. The distribution leaves room for an
    outflow along the polar axis and is derived from the
    model of Beckert \& Duschl \cite{Beckert04}.
    The spatial scale is in units of the dust sublimation radius.
    The mean number of clouds along a line of sight to the center drops
    below 1 for angels larger that $40^\circ$ from the midplane ($Z=0$).}
\label{density}
\end{figure}

Like in ordinary accretion disks, the effective viscosity from
(\ref{eq:visco}) allows mass to be accreted towards the black hole.
The conservation law of angular momentum for the torus clouds 
with a viscous torque can be integrated once for stationary accretion
\begin{equation}\label{angMomentum} 
   \nu \Sigma R^3 \frac{\partial \Omega}{\partial r} - \nu_i \Sigma_i R_i^3  
  \left.\left( \frac{\partial \Omega}{\partial r}\right) \right|_{r=R_i} = 
  - \frac{\dot{M}}{2 \pi} R^2 \Omega
  \left(1 - \frac{\Omega_i}{ \Omega} \left( \frac{R_i}{R}\right)^2 \right) \,.
\end{equation}
Special attention must be paid to the inner boundary at $R_i$ (index $i$).
This boundary will be at the sublimation radius for dust ($R\sim 1$\,pc), where
neither torque nor shear will vanish. In addition 
the torque at the inner boundary is most likely not well described  
by the viscosity (\ref{eq:visco}).  Here $\dot{M}$ is the total 
mass accretion rate,
which we assume to be constant at all radii throughout the torus.
The outer boundary of our model is determined by the feeding process.
The mass supply can be either a steady inflow driven by 
circumnuclear starformation,
discrete interactions with GMCs coming from large radii, or an inflow 
created by bars in the galactic potential.

In cloud-cloud collisions a fraction $\frac{1}{2}(1- \epsilon^2)$ of the 
average relative kinetic energy of clouds is dissipated, where $\epsilon$ is 
the coefficient of restitution, which approaches 1 for elastic collisions.
We use $\epsilon$ to parameterize cloud collisions and ignore 
the possibly more complex momentum redistribution in actual collisions. 
The energy dissipation 
appears as a sink term in the energy balance. 
For thin accretion disks with $\tau \sim 1$ we find $\epsilon >0.8$, 
which is  uncomfortably large. Smaller values of $\epsilon$ are possible 
in geometrically thick tori, because energy advection can balance 
the increased dissipation in collisions.
For a given set of parameters $(\epsilon, \dot{M}, M(R))$ 
the surprising result is that tori with
high accretion rates are geometrically thick 
$
H/R \sim \sqrt{2 R \dot{M}/( c_s\,M(R))}
$
in spite of the smaller $\epsilon$  of collisions. 
The elasticity, which leads to reasonably thick tori 
have $\epsilon$ in the range 0.2--0.6, which
might be provided by magnetic fields in the clouds.

\section{The Torus in  {NGC 1068} \label{N1068}}
For a detailed model of the  torus one needs to know 
the mass distribution in the center. For the case of NGC~1068
we collected the data in \cite{Beckert04}.
Some of the observed H$_2$O-maser spots
\cite{Greenhill97} around the radio core component 
S1 \cite{Gallimore04}
trace a rotating disk or ring. The maser velocities appear inconsistent with
Keplerian rotation around a central point mass. 
The rotation profile can be due to a massive disk
\cite{Hure02}, but we follow \cite{Kumar99} and find a 
model with a black hole  mass of $1.2\,10^7 M_\odot$
and a strongly concentrated  stellar cluster with $\rho_\star 
\propto R^{-\alpha}$ at large radii,
a core radius of $0.32$~pc and a core density
of $\rho_{\star,c} = 5.25\,10^6 M_\odot$\,pc$^{-3}$. 
Core radius and density are comparable
to the nuclear stellar cluster in the center of our 
Galaxy \cite{Schodel03}. 
\begin{figure}
\begin{center}
\includegraphics[width=.85\textwidth]{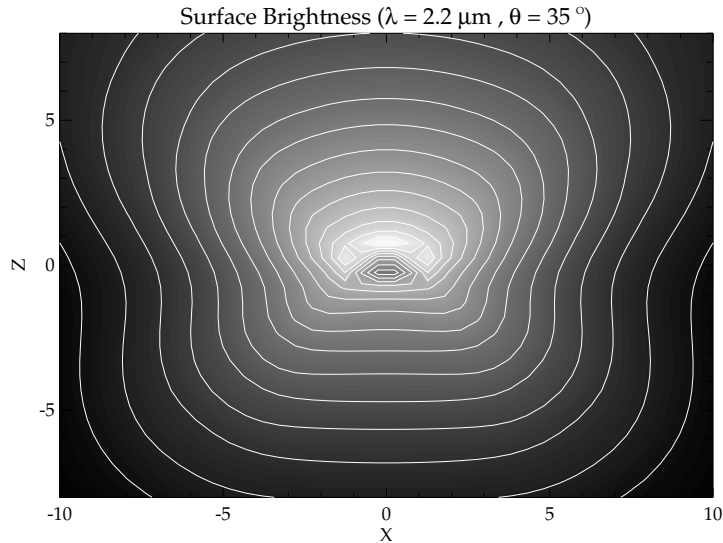}
\end{center}
\caption[]{
    K-band brightness distribution of our radiative transfer
  calculations based on the method of Nenkova et al. \cite{Nenkova02}
  and the scenario of Beckert \& Duschl \cite{Beckert04} for an
  inclination of $35^\circ$ from the midplane.
  The spatial scale is in units of the dust sublimation radius.
  The contour scale of the surface brightness is logarithmic 
  with a dynamic range
  of $2^{14}$.}
\label{k_band}
\end{figure}
The observational data from $0.3''$ to $4''$ from the nucleus are consistent 
with this mass distribution.
The central cluster extends out
to $250$~pc and the dusty torus lives in the gravitational potential 
of this cluster.

With the mass distribution, the model for individual clouds, 
and the cloud density distribution from Fig.~1 we performed 
radiative transfer calculations based on the method 
of Nenkova et al. \cite{Nenkova02}. An example of the 
surface brightness distribution is shown in Fig.~2. 
The simulation demonstrate that the observed  
core component in Speckle images \cite{Weigelt04} 
in the NIR indeed measures the size of the sublimation radius to 
$R = 1\pm 0.4$\,pc.

\section{Conclusions}
We described a dynamical approach to the radial and vertical
structure of a clumpy torus around an AGN following the ideas
of Krolik \& Begelman \cite{Krolik88} and Vollmer et al. \cite{Vollmer04}.
The vertical thickness is derived from an accretion
scenario for a geometrically thick torus, which is based on 
cloud-cloud collisions. The inferred mass accretion rates 
for scale heights $H > 0.5 R$ in the 
torus are larger than the Eddington rate for the central black hole 
and the model requires massive outflows from within 
the sublimation radius for dust. 
It is found that along typical lines of sight to
the center (for angles $< 45^\circ$ w.r.t. the torus midplane) 
between one and ten clouds on average obscure the AGN. 
Only a small fraction of the mass accreted through the torus
eventually reaches the black hole
but is sufficient to generate the ionizing luminosity.
An in depth 
analysis of cloud collisions in the environment of an AGN is required 
to test our model assumptions for $\epsilon$ 
and the structure of individual clouds.

%

\end{document}